\documentclass[twocolumn]{aastex631}

\usepackage{soul}
\usepackage{xcolor,cancel}
\usepackage{amsmath}

\begin{document}

\title{Truncated star formation and ram pressure stripping in the Coma Cluster}

\author[0009-0006-0440-2228]{Ariel O. Broderick}
\affiliation{Department of Physics \& Astronomy, University of Waterloo, Waterloo, ON N2L 3G1, Canada}

\correspondingauthor{Ian D. Roberts}
\email{ianr@uwaterloo.ca}

\author[0000-0002-0692-0911]{Ian D. Roberts}
\affiliation{Department of Physics \& Astronomy, University of Waterloo, Waterloo, ON N2L 3G1, Canada}
\affiliation{Waterloo Centre for Astrophysics, University of Waterloo, 200 University Ave W, Waterloo, ON N2L 3G1, Canada}
\affiliation{Leiden Observatory, Leiden University, PO Box 9513, 2300 RA Leiden, The Netherlands}

\author[0000-0002-1437-3786]{Michael J. Hudson}
\affiliation{Department of Physics \& Astronomy, University of Waterloo, Waterloo, ON N2L 3G1, Canada}
\affiliation{Waterloo Centre for Astrophysics, University of Waterloo, 200 University Ave W, Waterloo, ON N2L 3G1, Canada}
\affiliation{Perimeter Institute for Theoretical Physics, 31 Caroline St North, Waterloo, ON N2L 2Y5, Canada}



\begin{abstract}

We use 45 galaxies from the Mapping Nearby Galaxies at Apache Point Observatory survey to study the physical drivers of star formation quenching in the Coma cluster.  We measure specific star formation rate (sSFR) radial profiles for the Coma sample as well as a control sample of non-cluster field galaxies. We find that compared to the control sample, galaxies within the Coma Cluster have sSFR profiles that fall off more steeply with galactocentric radius. We then apply a toy model based on slow-then-rapid quenching via ram pressure stripping. We find that this model is able to reproduce the difference in sSFR profiles between field and Coma galaxies.  These results demonstrate that ram pressure stripping plays a significant role in quenching star formation in the nearest massive galaxy cluster.

\end{abstract}



\section{Introduction} \label{sec:intro}
The evolution of galaxies in clusters differ greatly from those that are relatively isolated in the field. It's been observed that clusters are dominated by red, quiescent galaxies, whilst sparser environments contain galaxies that are bluer with active star formation \citep[e.g.][]{Oemler_1974, Davis&Geller_1976, Butcher&Oemler_1978, Dressler_1980, Postman&Geller_1984}. This is due to the fact that galaxies in clusters are subjected to a host of different interactions, including gravitational interactions between all the surrounding galaxies, and hydrodynamical interactions from the act of moving through the intracluster medium (ICM). Such interactions include tidal effects which can remove the gas and stellar mass of a galaxy \citep[e.g.][]{Mayer_et_al_2006, Chung_et_al_2007}, harassment which is high speed fly-by encounters between galaxies that strips some gas \citep[e.g][]{Moore_et_al_1996}, viscous stripping \citep{Nulsen_1982}, thermal evaporation due to the hot ICM \citep{Cowie&Songaila_1977}, starvation which is the prevention of the accretion of more gas due to the hot ICM \citep[e.g.][]{Larson_et_al_1980, Peng_et_al_2015}, and ram pressure stripping \citep{Gunn&Gott_1972}.
\par
Ram pressure stripping (RPS) occurs as a galaxy moves through the ICM. Through this relative motion, a pressure is built up which can cause some of the intersteller medium (ISM) of the galaxy to be stripped away. Given that gas located in the outskirts of the disk is less-strongly bound than gas near the galaxy center, this stripping occurs from the outside-in. In rich clusters, ram pressure is thought to play the dominant role in quenching galaxies. Evidence for this includes the high \textsc{HI} deficiencies of cluster galaxies relating to their orbits \citep[e.g.][]{Vollmer_et_al_2001, Boselli&Gavazzi_2006, Gavazzi_et_al_2013, Boselli_et_al_2014, OmanBaheHealy2021}, the presence of one-sided tails extending beyond the galaxy disk (so-called ``jellyfish'' galaxies e.g. \citealt[][]{Gavazzi_et_al_2001, SmithLuceyHammer2010, poggianti2017, boselli2018, Jaffe_et_al_2018, roberts2021_LOFARclust, roberts2021_LOFARgrp}), metallicities being higher closer to the centre of the cluster meaning a denser ICM \citep[e.g.][]{Maier_et_al_2019, Ciocan_et_al_2020}, more quenched galaxies being found in regions of higher ICM density \citep{Roberts_et_al_2019}, and outside-in stripping of cold gas in cluster galaxies \citep[e.g.][]{Cortese_et_al_2021, Zabel_et_al_2022}.
\par
Furthermore, the outside-in nature of quenching via RPS can be probed by exploring radial profiles of star formation in cluster galaxies. For other galaxies, like the Virgo cluster, radial profiles showed that star formation rates are concentrated near the center of the galaxies \citep[e.g.][]{Koopmann&Kenney_2004, Koopmann_et_al_2006}. \citet{Schaefer_et_al_2017} found that this was also true when studying the radial profiles of cluster galaxies, indicating that galaxies in clusters may be experiencing outside-in quenching. There have been numerous studies using radial profiles to better understand quenching of galaxies, such as seeing whether it depends on environment density or if stellar mass has any affect \citep[e.g.][]{Spindler_et_al_2018, Schaefer_et_al_2019, Coenda_et_al_2019, OwersHudsonOman2019}.
\par
In this work we use resolved optical spectroscopy from the Mapping Nearby Galaxies at Apache Point Observatory (MaNGA, \citealt{Westfall_et_al_2019, Belfiore_et_al_2019}) survey to explore the impact of RPS on star formation within galaxies in the Coma Cluster. As the nearest massive cluster, Coma is an excellent laboratory for improving our understanding of galaxy evolution in dense environments. Its proximity permits resolved studies of star formation at kiloparsec scales in member galaxies. Additionally, many galaxies within the Coma cluster are $\textsc{HI}$ deficient \citep[e.g.][]{Gavazzi_1987, Gavazzi_1989, Molnar_et_al_2022}, already suggesting a significant role played by RPS.
\par
The outline of the paper is as follows. In Section~\ref{sec:data} we describe the sample selection, the production of stellar mass and SFR maps, and the method used for calculating radial profiles of star formation. In Section~\ref{sec:results} we present the main results from this work, including observed radial profiles for Coma galaxies and field galaxies, as well as the application of a simple model of RPS to these observed trends. In Sections \ref{sec:discuss} and \ref{sec:conclus} we discuss these results and present our conclusions. Throughout we have assumed a flat $\mathrm{\Lambda CDM}$ cosmology with $\Omega_m = 0.3$, $\Omega_\Lambda = 0.7$, and $H_0 = 70\,\mathrm{km\,s^{-1}\,Mpc^{-1}}$.

\section{Data and Methods} \label{sec:data}

\begin{figure*}[!ht]
\centering
\includegraphics[width = \textwidth]{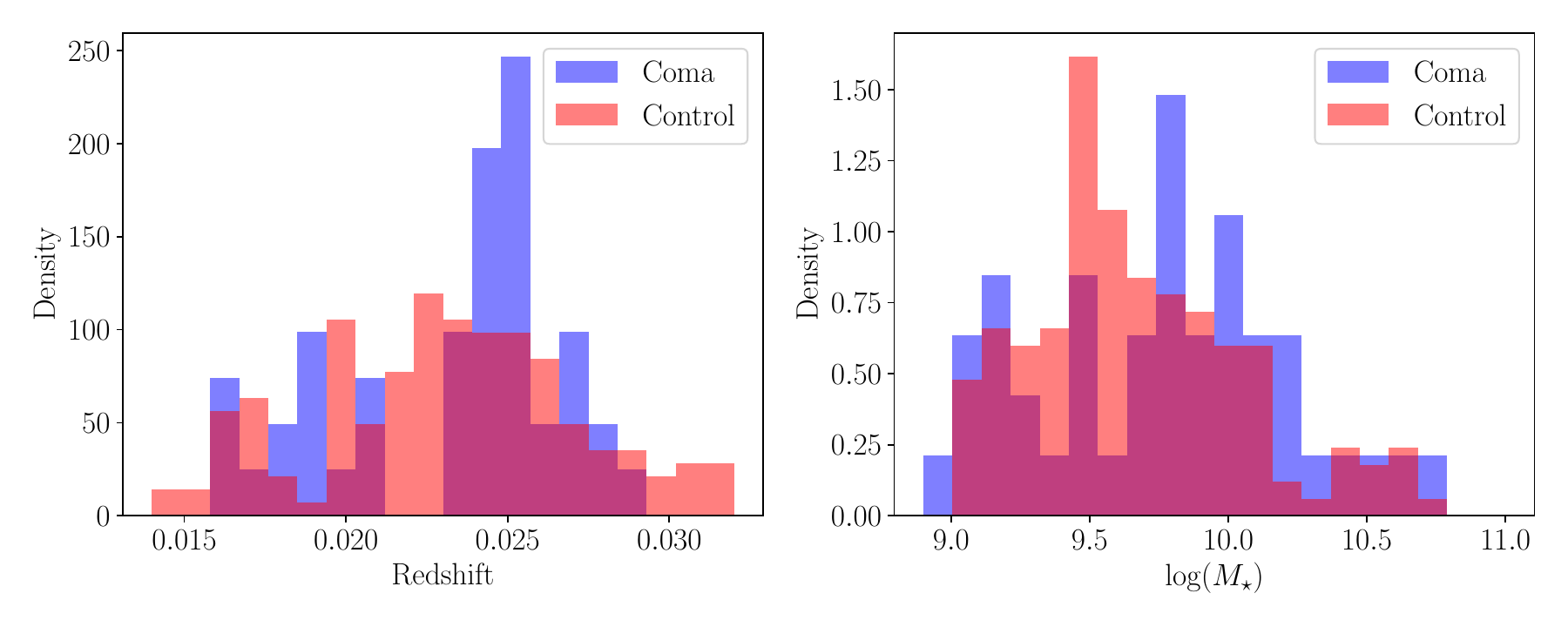}
\caption{Redshift and logged integrated stellar mass distributions of Coma galaxies (blue) and the matched control galaxies (red).}
\label{fig:M_z_distributions}
\end{figure*}


\subsection{Galaxy samples} \label{sec:galaxy_samples}

We construct a sample of satellite galaxies in the Coma Cluster that have publicly available integral field unit (IFU) spectroscopy from the MaNGA survey. We initially select all star-forming and green-valley galaxies from the final MaNGA sample that are within $R_{200}$ of Coma (in projection) and have projected velocity offsets that are $<3000\,\mathrm{km\,s^{-1}}$ from the cluster centroid. Our star formation selection is based on a single cut in specific star formation rate at $\mathrm{\log (sSFR / yr^{-1})} > -11.7$. This selection gives 106 star-forming and green-valley member galaxies in Coma with MaNGA spectroscopy. This initial selection is based on integrated SFRs and stellar masses from SED fitting \citep{salim2016,salim2018}. However, further inspection revealed that there were a number of very low mass Coma galaxies that passed this initial selection, but actually had no detected $\mathrm{H\alpha}$ emission in their MaNGA maps. This could simply be due to large uncertainties on the SED SFRs, or alternatively it could be that these galaxies were quenched very recently. Following the spirit of our initial star formation selection, and because we aim to measure resolved properties of $\mathrm{H\alpha}$ emission in Coma galaxies, we cull any galaxies from the sample without MaNGA-detected $\mathrm{H\alpha}$ emission. This leaves a final Coma sample of 45 galaxies.   
\par
\begin{figure}[!ht]
\centering
\includegraphics[width = \columnwidth]{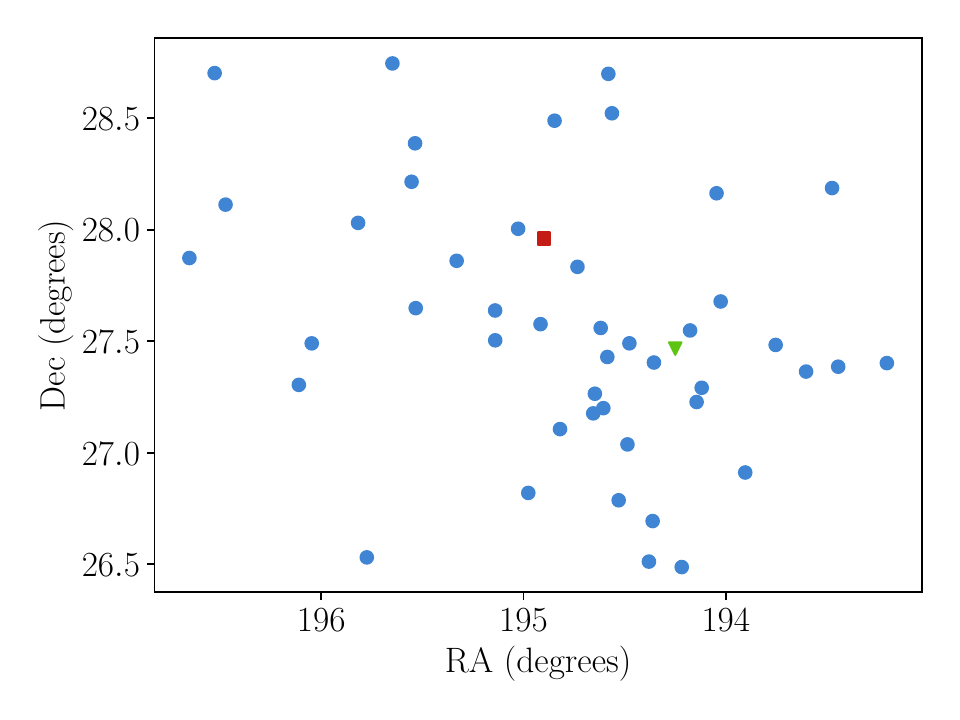}
\caption{Blue dots indicate the position of the Coma galaxies in our sample. The red square is the center of the cluster and the green triangle indicates the position of NGC 4839}
\label{fig:Coma_Map}
\end{figure}

In Fig.~\ref{fig:Coma_Map} we show the on-sky positions of our Coma sample alongside the center of the Coma Cluster (red square) and the position of NGC4839 (green triangle), the central galaxy in an infalling group. We note that our galaxies are well distributed across the full Coma Cluster, with no obvious overdensity at the location of NGC4839. Thus we expect any observed trends to be primarily driven by the large-scale cluster environment.
\par
We also compile a control sample of field galaxies that will be used as a comparison for the cluster sample. To construct the control sample, we first selected all galaxies from the \citet{lim2017} group catalogue that are isolated (i.e.\ part of a single-member group) or part of a group with halo mass $<10^{12.5}\,\mathrm{M_\odot}$ and within $3000\,\mathrm{km\,s^{-1}}$ of the Coma Cluster redshift. We then cross-match this subset with the MaNGA final data release sample to select those galaxies with IFU spectroscopy.
\par
Lastly, to ensure that there are not stellar mass biases between the Coma and control samples, we selected a subsample of control galaxies that are well-matched to the stellar mass distribution of the Coma sample. We have run a two-sample Anderson-Darling test between the stellar mass distributions for the Coma and control samples. This test outputted a $p$-value of 0.119, indicating that there is no statistical evidence they were drawn from different parent distributions. 

\subsection{Stellar mass and SFR maps} \label{sec:M_SFR_maps}

For all galaxies we make use of resolved stellar mass and star formation rate (SFR) maps. We use dust-corrected stellar mass maps from the Pipe3D value-added catalogs \citep{Sanchez_et_al_2016a,Sanchez_et_al_2016b}. For each stellar mass map we apply the ``dezonification'' map in order to reduce the signature of the spatial binning (see \citealt{cidfernandes2013} for more details). 
\par
We compute SFR maps based off of the extinction-corrected $\mathrm{H\alpha}$ luminosity. We first mask all spaxels with $\mathrm{S/N} < 3$ in either the $\mathrm{H\alpha}$ or $\mathrm{H\beta}$ emission lines. We correct the $\mathrm{H\alpha}$ luminosities for dust extinction using the Balmer decrement. The extinction-corrected $\mathrm{H\alpha}$ luminosity is then converted to a SFR using the calibration from \citet{kennicutt2012}. Lastly, in the SFR maps we mask any spaxels that are classified as AGN according to the BPT diagram using \textsc{NII}/H$\alpha$, \textsc{SII}/H$\alpha$, and \textsc{OIII}/H$\beta$ lines and a cutoff of S/N $< 3$ \citep{Kewley_et_al_2006}.

\subsection{Star formation radial profiles} \label{sec:ssfr_profiles}

\begin{figure}[!ht]
\centering
\includegraphics[width=\columnwidth]{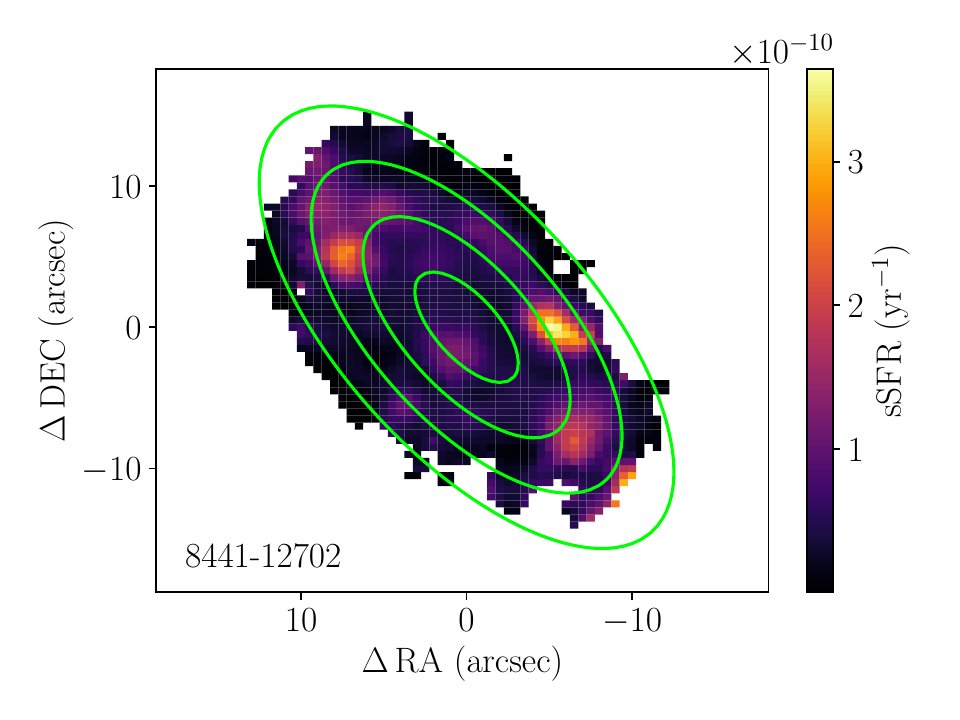}
\includegraphics[width=\columnwidth]{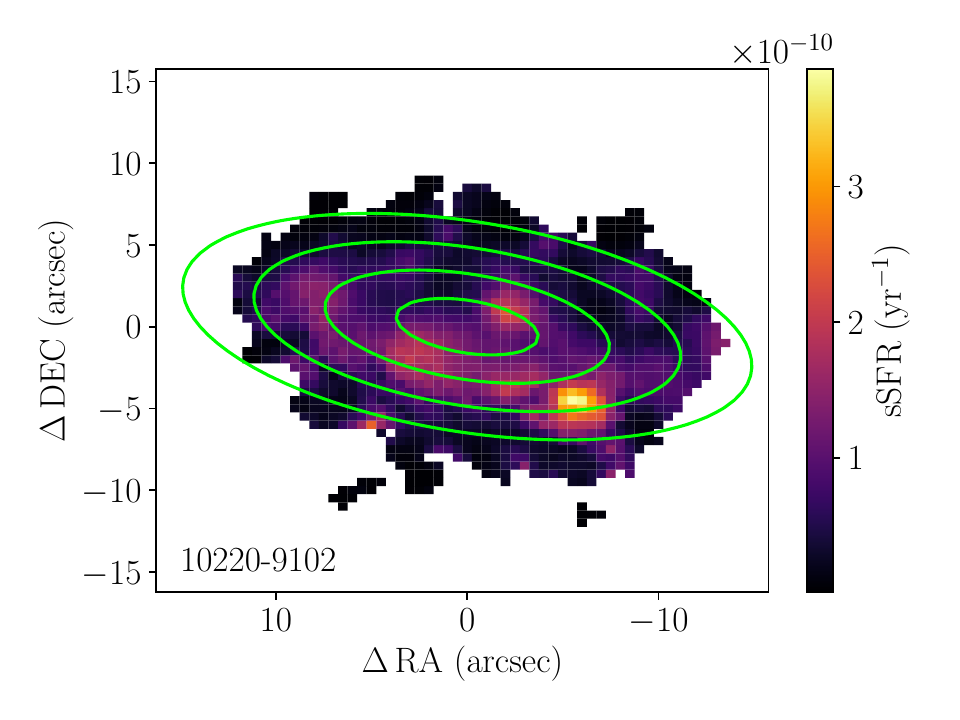}
\includegraphics[width=\columnwidth]{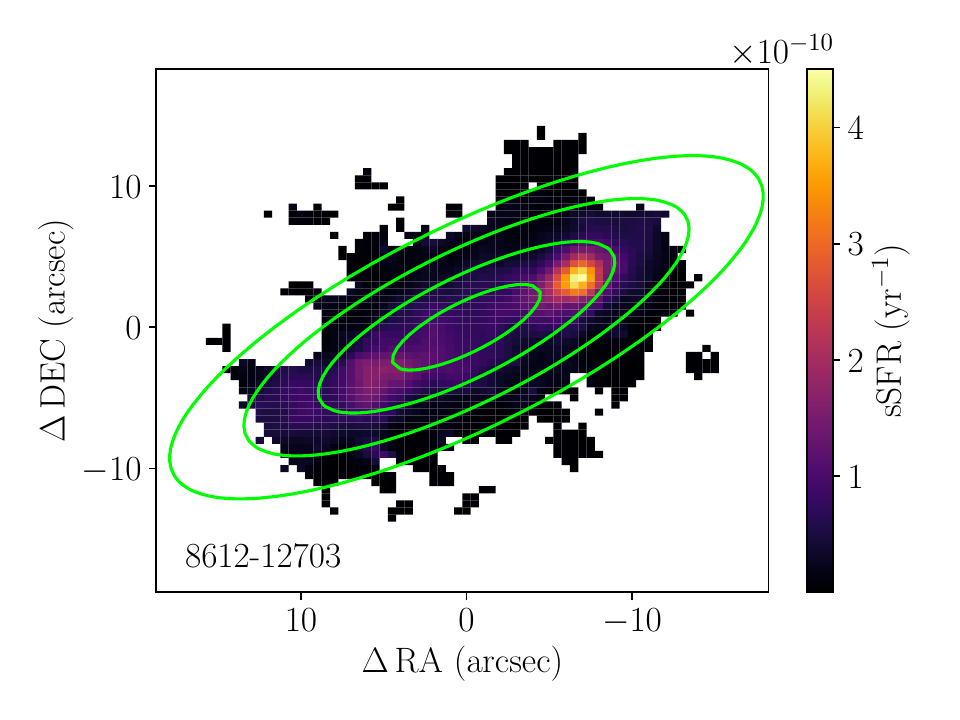}
\caption{sSFR maps. Green rings are bin edges ($[0,0.5,1,1.5,2]R_e$). These 3 galaxies are from the control sample}
\label{fig:Bin_Diagrams_1}
\end{figure}

To compute star formation radial profiles, we take each galaxy and sort the star forming pixels into elliptical radial bins (bin edges are $[0,0.5,1,1.5, 2]\,R_e$). The average sSFR in each annulus is calculated by dividing the total SFR (which we get by summing the SFR of all pixels in the annulus) by the total stellar mass in the annulus. This is done for every Coma and control galaxy. For each radial bin, the mean and standard error of the sSFR are calculated in order to give the average radial profile and its spread.

\section{Results} \label{sec:results}

\subsection{Observed Trends}


In Fig.~\ref{fig:sSFR_Model} we show sSFR radial profiles for galaxies in the Coma and control samples. In the innermost bin the sSFR between the two samples are similar. However, with increasing galactocentric radius the difference between the Coma and control samples increases, with the Coma sSFR profiles falling off more steeply with radius. Qualitatively, this difference between the shape of the Coma and control radial profiles is consistent with expectations from RPS, where ram pressure can more easily strip gas (and thus quench star formation) from the galaxy outskirts. In the next section, we quantitatively confirm this observation via a toy model of RPS.

\subsection{Quenching Model}

\begin{figure}[!ht]
\includegraphics[width = \columnwidth]{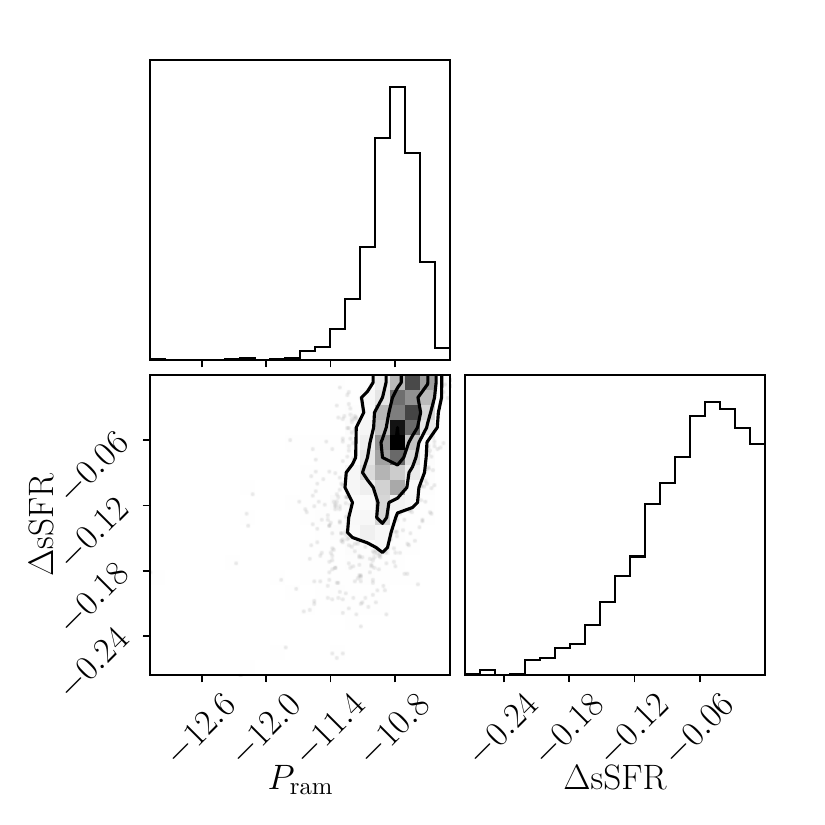}
\caption{Corner plot for the ram pressure used to match to the Coma galaxies ($P_\mathrm{{ram}}$, in units of $\mathrm{g\, cm^{-1} \, s^{-2}}$) and the additive shift in log space ($\mathrm{ \Delta sSFR}$).
\label{fig:Corner_Plot}}
\end{figure}

\begin{figure}[!ht]
\centering
\includegraphics[width=\columnwidth]{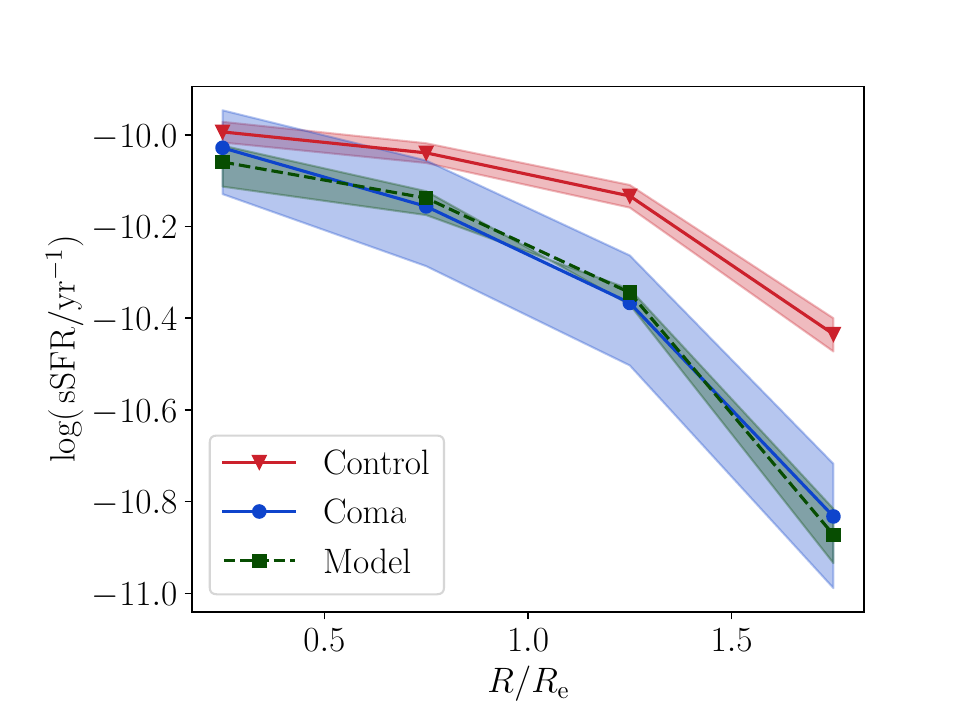}
\caption{Each point (for the Coma and Control) is the mean of all the average sSFR of the bin for that group. The points for each bin are placed midway between each bin edge on the $x$-axis (so the mean for the data between 0 and $0.5R_{\rm e}$ has an $x$ value of $0.25R_{\rm e}$). Shaded area is standard deviation of the average sSFR in each bin. After applying the ram pressure model to the control, we get the dark line which is the 50th percentile. The shaded region for the model are from the 16th to the 84th percentile.}
\label{fig:sSFR_Model}
\end{figure}

We now test whether the differences in sSFR radial profiles shown in Fig.~\ref{fig:sSFR_Model} can be accounted for using a simple toy model of quenching via RPS. Our model is motivated by the slow-then-rapid framework for satellite quenching that has been advocated by a number of previous works \citep[e.g.][]{Wetzel_et_al_2013,Roberts_et_al_2019,Maier_et_al_2019b, ReevesHudsonOman2023}. Our approach is to define a toy quenching model that we can apply to the observed sSFR maps for our control sample galaxies. We then compute radial profiles from the control sample sSFR maps after being processed through the quenching model, and compare to the observed radial profiles for the Coma galaxies.
\par
The quenching model has two components. First, an additive (in log space) sSFR offset term, $\mathrm{\Delta sSFR}$, which is constant and applied to all pixels in the sSFR map. This can be considered the ``slow'' component and can be interpreted physically as a contribution from pre-processing, gas consumption (starvation), or both \citep[e.g.][]{Roberts_et_al_2019}. It is also possible that the existence of backsplash galaxies in the Coma sample may contribute to this difference. The second, ``rapid'' component is tied explicitly to RPS. We take a standard approach \citep{Gunn&Gott_1972, Rasmussen_2006, Roberts_et_al_2019} and compute whether each pixel in the sSFR map is susceptible to RPS. This occurs where:
\begin{equation}
    P_{\rm ram} > 2(\Sigma_{\star} + \Sigma_{\rm gas})\Sigma_{\rm gas}
\end{equation}
In this case, $\mathrm{\Sigma_{gas}}$ refers to the sum of the atomic and molecular gas (i.e. $\mathrm{\Sigma_{gas}} = \mathrm{\Sigma_{HI}} + \mathrm{\Sigma_{H_2}}$). If a pixel satisfies this inequality then the pixel is ``stripped'' and its sSFR is set to 0. $\Sigma_{\star}$ is taken from the stellar mass maps described in Section~\ref{sec:M_SFR_maps}. To estimate $\Sigma_{\rm H_2}$ on a pixel-by-pixel basis we use a standard Kennicutt-Schmidt relation from \citet{Bigiel_et_al_2008}:
\begin{equation}
    \log \left(\frac{\Sigma_\mathrm{SFR}}{\mathrm{M_{\odot}\,yr^{-1}\,kpc^{-2}}}\right) = 1.01\log \left(\frac{\Sigma_\mathrm{gas}}{10\,\mathrm{M_{\odot}\,pc^{-2}}}\right) - 2.12,
\end{equation}
\noindent
given $\Sigma_\mathrm{SFR}$ from the SFR maps described in Section~\ref{sec:M_SFR_maps}. 

\begin{figure}[!ht]
\centering 
\includegraphics[width=\columnwidth]{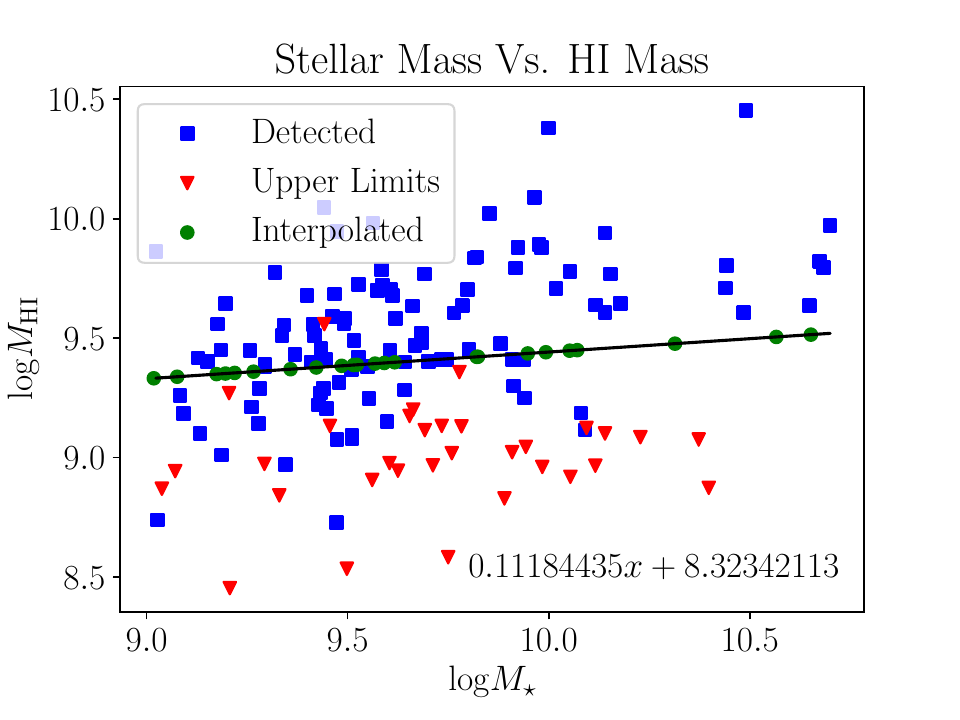}
\caption{Logged stellar mass vs logged \textsc{HI} mass. The blue points belong to control galaxies with detected \textsc{HI} masses in the \textsc{HI} MaNGA survey. The red points represent control galaxies that had only upper limits for their \textsc{HI} mass. The black line is the line of best fit to the data. The green points represent galaxies that were not part of the \textsc{HI} MaNGA survey and had to be interpolated along the line of best fit.}
\label{fig:HI_Vs_Stellar}
\end{figure}

To get the \textsc{HI} mass for each pixel for each galaxy in our control sample, we use the \textsc{HI} masses from the \textsc{HI} MaNGA survey. If a galaxy only had an upper limit, that upper limit was used in place of its \textsc{HI} mass, i.e.\ we assume that these galaxies have as much \textsc{HI} as possible. Some of the galaxies in our control sample were not part of the survey so we interpolated their \textsc{HI} masses by assuming a linear relationship between stellar mass and \textsc{HI} mass. We then fit a line to the data we did have (using the data with only limits as censored points), and interpolated the \textsc{HI} of those galaxies along the fitted line. This best-fit is shown in Fig.~\ref{fig:HI_Vs_Stellar}. We then assume an axisymmetric distribution and use the following equation:
\begin{equation}
    \Sigma_{\rm HI}(r) = \mathrm{\frac{M_{HI}}{2\pi R_d^2}}e^{-\frac{r}{\rm R_d}}
\end{equation}
Where $\mathrm{R_d} = \frac{1.7 \rm R_e}{2}$ is the approximate radius of the \textsc{HI} disk. We add $\Sigma_{H_2}$ and $\Sigma_{HI}$ to get $\Sigma_{\rm gas}$.
\par

We apply this model to the galaxies in the control sample, calculate radial sSFR profiles for model-processed maps, and then determine best-fit $\Delta \mathrm{sSFR}$ and $P_\mathrm{ram}$ in order to reproduce the sSFR radial profiles for the  Coma sample. Fitting is done using the \texttt{Python} Markov Chain Monte Carlo package, \texttt{emcee} \citep{Foreman-Mackey_et_al_2013}.
\par
In Fig.~\ref{fig:Corner_Plot} we show the fitting results for the  two free parameters from our quenching model, $P_{\rm ram}$ and $\Delta \mathrm{sSFR}$, with all parameters well constrained. In Fig.~\ref{fig:sSFR_Model} we show the best-fit radial profile resulting from the application of our quenching model to our control galaxy sample (dashed lines), compared to the observed radial profile shown in Section~\ref{sec:ssfr_profiles} (solid lines). The Coma trend is well reproduced by the model, with best-fit ram pressure strengths of $1.5\,_{-0.6}^{+0.9} \times 10^{-11}\,\mathrm{g\,cm^{-1}\,s^{-2}}$ and a $\Delta \mathrm{sSFR}$ of $-0.06_{-0.06}^{+0.04}$.
\par
We note that the best-fit value for $\mathrm{\Delta sSFR}$ is consistent with zero. At face value this indicates that for our sample of Coma galaxies there is little need for the ``slow-quenching'' component of our model, and that the difference between the star formation profiles of Coma and field galaxies can be explained purely with the addition of ram pressure stripping. Though we argue that correspondence for the inner-most sSFRs between the Coma and control samples is more complicated than this, and driven both by the fact that ram pressure can reduce star formation in galaxies but also can temporarily enhance star formation during the early stages of the process \citep[e.g.][]{Vulcani_2018, Roberts&Parker_2020}. We discuss this point in more detail in Sect.~\ref{sec:discuss}.
\par
Finally, we have performed this full analysis using the median as our estimator instead of the mean and we show this in Appendix~\ref{sec:median}. This complementary analysis provides qualitatively similar results (with larger uncertainties) as well as quantitatively a best-fit value for $P_\mathrm{ram}$ that is fully consistent with the value derived in the main body of the paper.

\section{Discussion} \label{sec:discuss}

\subsection{Derived ram pressure strengths}

\begin{figure}[!ht]
\includegraphics[width = \columnwidth]{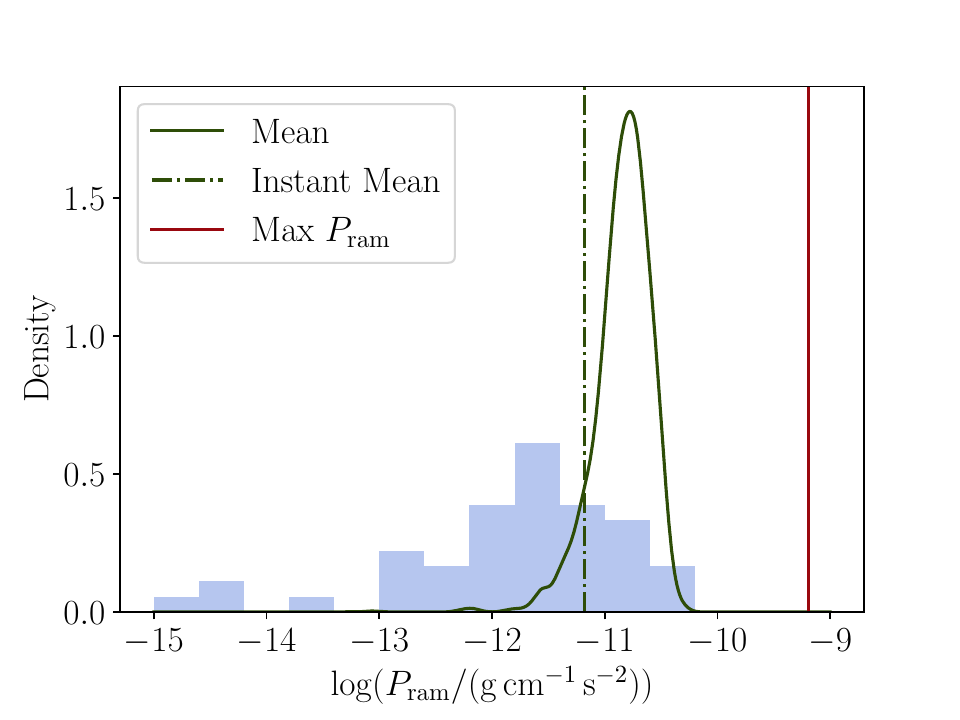}
\caption{Logged ram pressure distribution of the Coma galaxies. The solid line is the fitted ram pressure distribution from the model . The dashed-dotted line is the calculated logged mean of the ram pressures for the  Coma group. The solid vertical line is the maximum ram pressure we expect a galaxy in the Coma cluster could experience.
\label{fig:Pram_Hist}}
\end{figure}

In the previous section, we use our toy model to derive the ram pressure strength required to reproduce the observed sSFR radial profile for the Coma sample. An important model validation is to check that these derived values are broadly consistent with the expected ram pressure strengths experienced by satellite galaxies in Coma. We can roughly estimate the ram pressure for the Coma galaxies using the following equations:
\begin{equation}
    P_{\rm ram}(r) = \rho_{\rm ICM}(r)\,v^2 
\end{equation}
\noindent
where,
\begin{equation}
    \rho_\mathrm{ICM}(r) = \rho_\mathrm{ICM}(0)\left(1+\left(\frac{r}{r_c}\right)^2\right)^{-\frac{3\beta}{2}}
\end{equation}
\noindent
and,
\begin{equation}
    v = 3 \times 10^{10}\,|z_{\rm galaxy}-z_{\rm Coma}| \quad \quad \mathrm{[cm\,s^{-1}]}.
\end{equation}
We assume $z_{\rm Coma} = 0.0231$ and a beta-profile for the ICM density from \citet{Chen_et_al_2007} with $\rho_{ICM}(0) = 5 \times 10^{-27}\,\mathrm{g\,cm^{-3}}$, $r_c = 343\,\mathrm{kpc}$, and $\beta = 0.643$. We also apply factors of $\sqrt{3}$ and $\pi / 2$ \citep[e.g][]{Jaffe_et_al_2015} to the velocity offsets and clustercentric radii respectively, by multiplying to deproject these quantities, on average. Thus, given the projected phase space coordinates of a given galaxy in Coma, we can roughly estimate the local ram pressure to that galaxy via the above relations.
\par
In Fig.~\ref{fig:Pram_Hist} we show histograms corresponding to the local ram pressure estimated for all galaxies in the  Coma sample. With the solid line, we show the best-fit ram pressure strength distribution. The vertical, dash-dotted line is the log of the mean local ram pressure. For the Coma sample, the best-fit ram pressure strength is biased toward the upper quartile of the distributions for the galaxy-by-galaxy ram pressure estimates. Though, there is still substantial overlap between the two within the posterior distribution . We also calculate the maximum expected ram pressure as
\begin{equation}
    P_\mathrm{ram, max} = \rho_\mathrm{ICM}(0)\,v_\mathrm{max}^2
\end{equation}
\noindent
where $v_\mathrm{max}$ is the maximum velocity offset observed in our Coma sample of galaxies. This maximum ram pressure is shown as the solid vertical line in Fig.~\ref{fig:Pram_Hist}. As expected our best-fit ram pressure from the model is below this maximum value, since it is very unlikely that a given galaxy will have an orbit such that it experiences that maximum possible ram pressure strength.
\par
We note that this is a very rough comparison between model and data, and there are a number of implicit assumptions worth bearing in mind. The ram pressure strength derived from the model is the peak ram pressure, whereas the local ram pressure estimated on a galaxy-by-galaxy basis is an estimate of the instantaneous ram pressure. Thus if a significant number of galaxies in the Coma sample are post-peak ram pressure, then by construction the values derived from the model will be comparatively skewed towards high ram pressure strengths. Along these lines, the  Coma sample will, to some extent, be contaminated by backsplash galaxies that have already made a pericentric passage.  Finally, we apply average deprojection factors to the velocity offsets and clustercentric radii, but the applicability of these average factors on a galaxy-by-galaxy basis is highly uncertain.

\subsection{Implications for satellite quenching in Coma}

Broadly speaking, the observed sSFR radial profiles in Coma are consistent with quenching via RPS. This is consistent with the change in slope that we observe for the Coma sample.  This is in line with previous studies that have advocated for the strong role of RPS in Coma \citep[e.g.][]{Yagi_et_al_2010,Gavazzi_et_al_2018,Roberts&Parker_2020,Roberts_et_al_2024_coma_radio}. Indeed, 14 Coma galaxies in our sample are known to be jellyfish galaxies with observed ram pressure stripped tails in the radio continuum \citep{roberts2021_LOFARclust}. We note that the main qualitative result from this work, the steeper decline of star formation for Coma galaxies, persists even when we remove these radio jellyfish galaxies from our sample. This likely suggests that the non-jellyfish galaxies in our sample are at a later stage of ram pressure stripping without a clear stripped tail but still showing the signatures of outside-in quenching in the disk. Other wavelengths have also been used to identify galaxies undergoing RPS in the Coma cluster. Two of the galaxies in our sample were also identified as having $\mathrm{H\alpha}$ tails by \citet{Yagi_et_al_2010}, one was identified as having a UV tail by \citet{SmithLuceyHammer2010}, nine are part of the broadband-optical selected ram pressure candidates presented in \citet{Roberts&Parker_2020}, and three are part of the sample of galaxies with one-sided \textsc{Hi} tails in \citet{Molnar_et_al_2022}. This results in a total of 16 Coma galaxies in our sample being known jellyfish galaxies. In general, all of these jellyfish galaxies, regardless of how they were identified, tend to have central sSFRs that fall above the field relation.
\par
In this work we employ a model inspired by ``slow-then-rapid'' models of galaxy quenching, but as shown in Sect.~\ref{sec:results}, at face value the data do not seem to require a slow-quenching component. That said, when we use the median-estimator models (Appendix~\ref{sec:median}) the best-fit model does prefer a non-zero value of $\mathrm{\Delta sSFR} \approx -0.41$. One way to reconcile this is to note that the mean-estimator is more sensitive to anomalously high or low sSFRs, and also to note that jellyfish galaxies in Coma (likely at an early stage of RPS) are known to show enhanced SFRs thought to be driven by ram pressure compressed gas in the ISM. The anomalously high sSFRs for the jellyfish galaxies in our sample may be skewing the mean Coma radial profile shown in Fig.~\ref{fig:sSFR_Model} upwards toward the value for the control sample. This is confirmed by the fact that the average sSFR radial profile for these known jellyfish galaxies falls well above the control sample for the central radial bins (plot not shown). When we run our quenching model for Coma galaxies having removed all known jellyfish galaxies from the sample, we obtain a best fit value of red $\mathrm{\Delta sSFR = -0.56^{+0.18}_{-0.13}}$. This explanation is somewhat speculative, but also is consistent with all of the observations shown in this work. This highlights the complications that can arise when trying to interpret the effects of ram pressure which, depending on the situation, may decrease or increase observed star formation.
\par
Looking at radial profiles, \citet{Schaefer_et_al_2017} observed that star formation for cluster galaxies was centrally concentrated. This is also observed in the Virgo cluster \citep[e.g.][]{Koopmann&Kenney_2004, Koopmann_et_al_2006}. \citet{Cortese_et_al_2012} observes that in the Virgo cluster, the colour of the inner UV disk of \text{Hi} deficient galaxies is less affected than the colour of the outer UV disk when compared to \textsc{HI} normal galaxies, which is consistent with the outside-in quenching nature of RPS. \cite{OwersHudsonOman2019} found that, in high density cluster environments, galaxies with recently-quenched star formation in the outer disk had ongoing star formation in the inner regions. Our observations are consistent with these results. However, our results are not consistent with \citet{Schaefer_et_al_2019}, who observed that more specifically, only cluster galaxies with stellar masses greater than $10^{10}M_{\odot}$ from groups with a dark matter halo mass greater than $10^{12.5}M_\odot$ had centrally concentrated star formation that was likely due to outside-in quenching. Most of the galaxies in our Coma sample have stellar masses less than $10^{10}M_\odot$ and we still observe more centrally concentrated star formation, though we note that the massive Coma Cluster is a very different environment than the lower mass groups probed by \cite{Schaefer_et_al_2019}. \citet{Coenda_et_al_2019} found that there was evidence for outside-in quenching for galaxies in groups with stellar masses $9 \leq \log(\frac{M_\star}{M_\odot}) \leq 10$, but not for galaxies in groups with higher stellar masses. Our results seem to be consistent with the first part.

\section{Conclusion} \label{sec:conclus}

 We procured a sample of star forming and green valley Coma galaxies and then calculated the average sSFR in each radial bin for both the Coma galaxies as well as a control sample of field galaxies. Then we made a simple toy model of slow-then-rapid-quenching via RPS. We found that:

\begin{itemize}
    \item There is evidence of outside-in quenching within the Coma cluster. 

    \item These observations can be roughly reproduced by a toy model of slow-then-rapid quenching in the dense cluster environment. The slow component is consistent with gas depletion and/or pre-processing whereas the rapid component appears strongly tied to ram pressure stripping.
\end{itemize}

There are currently ${>}100$ galaxies in Coma with public IFU spectroscopy from MaNGA (only 45 were used in this work), though given the MaNGA size selection this sample primarily covers intermediate- and low-mass galaxies. We have been awarded time for the Coma Legacy IFU Survey (CLIFS) on the new WEAVE spectrograph. This survey will complete the high-mass IFU coverage in Coma (for star-forming and green valley galaxies), and ultimately has the goal of obtaining IFU spectroscopy (in conjunction with existing MaNGA data) for all star-forming and green-valley galaxies in Coma with $M_\star > 10^9\,\mathrm{M_\odot}$. As data for CLIFS are delivered, this will only increase the statistics for the sort of analysis presented in this work, and it turn will provide even tighter constraints on star formation histories and quenching for satellite galaxies in Coma.

\appendix

\section{Median estimator for radial profiles} \label{sec:median}

\begin{figure}[!ht]
\includegraphics[width = \columnwidth]{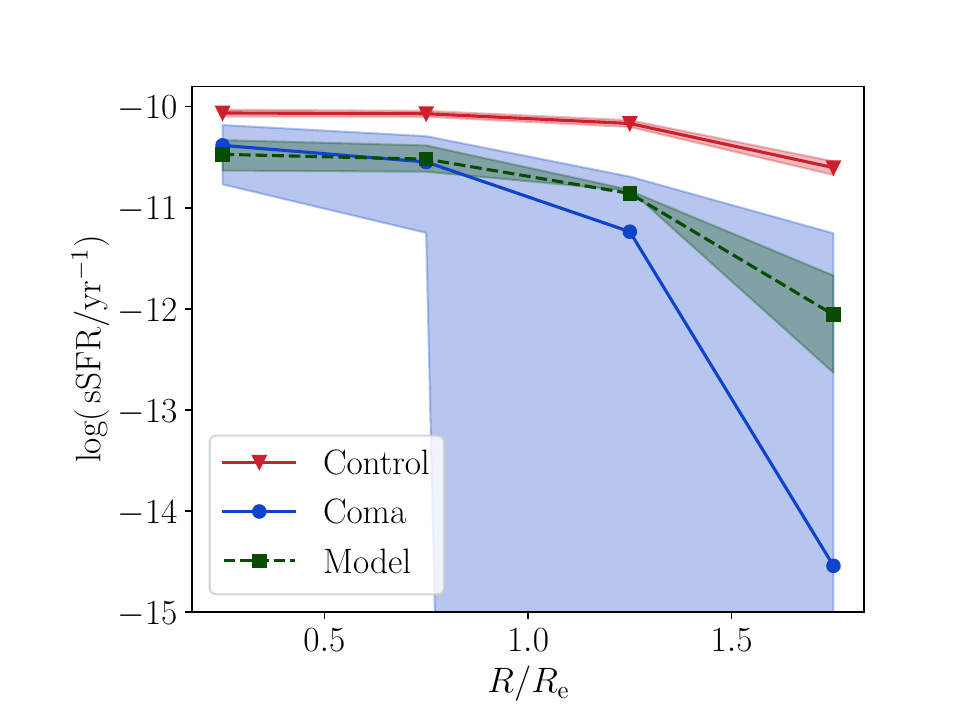}
\caption{Each point (for the Coma and control) is the median of all the average sSFR of the bin for that group. Points are placed midway between each bin edge on the $x$-axis. Shaded region is the standard error of the average sSFR in each bin. Just like Fig.~\ref{fig:sSFR_Model} except the model is fitted to the median sSFR radial profile.
\label{fig:Model_Median}}
\end{figure}

\begin{figure}[!ht]
\includegraphics[width = \columnwidth]{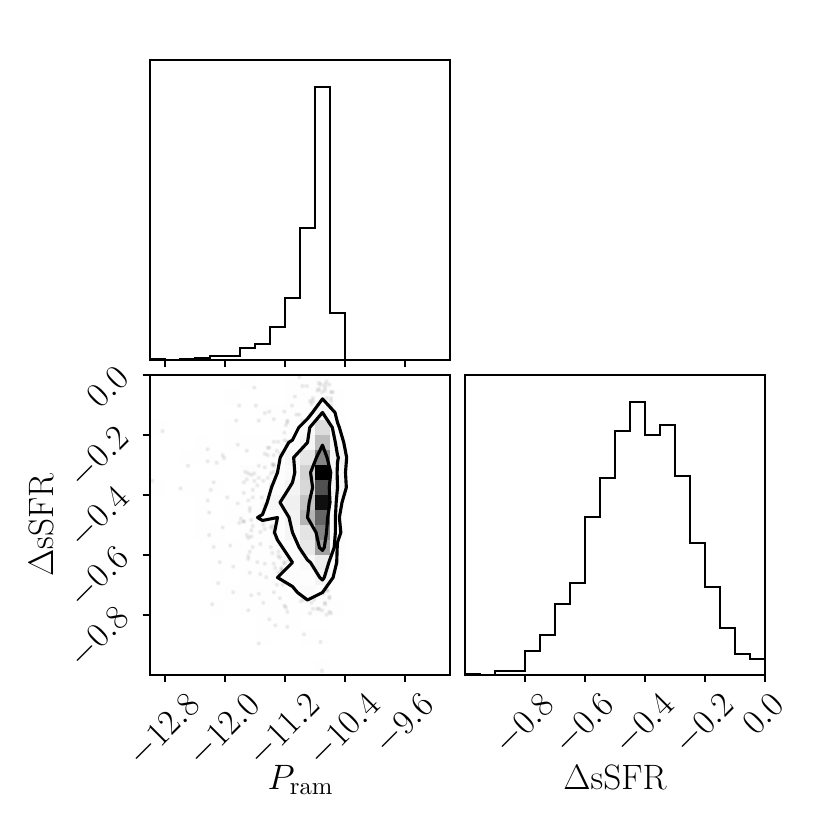}
\caption{Corner plot for ram pressures ($\mathrm{P_{ram}}$ is in units of $\mathrm{g\, cm^{-1}\, s^{-2}}$) fitted to the Coma radial profile and the additive shift in log space ($\mathrm{\Delta sSFR}$)
\label{fig:Corner_Plot_Median}}
\end{figure}

\begin{figure}[!ht]
\includegraphics[width = \columnwidth]{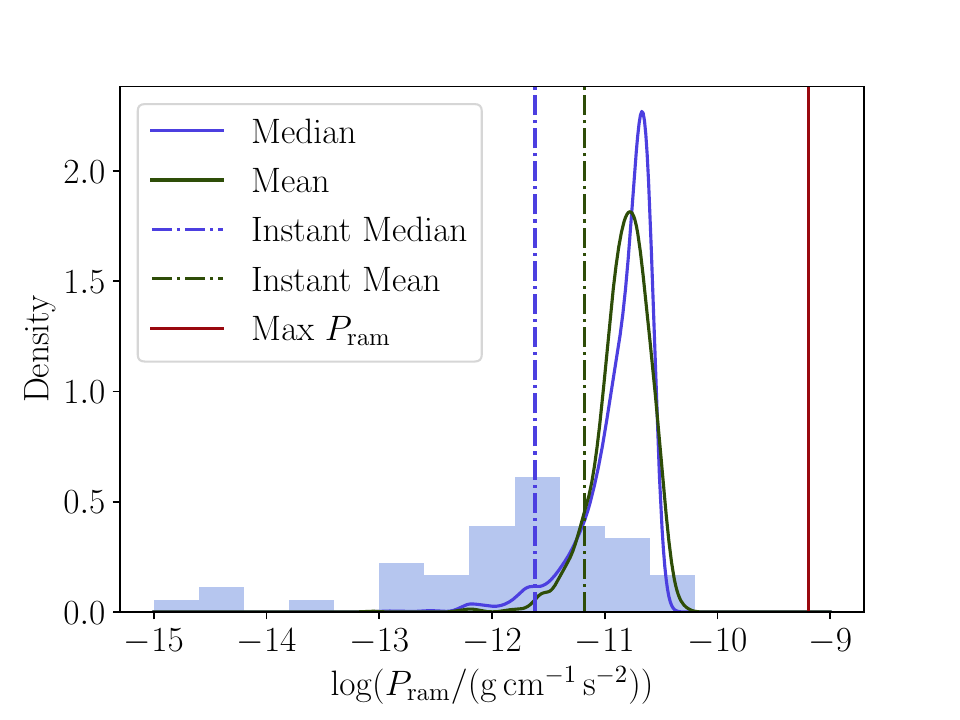}
\caption{Shaded region is the distribution of logged instantaneous ram pressures of the Coma galaxies. Solid lines represent the fitted ram pressure distribution, where median was fit to the median profile and the mean was fit to the mean profile. Dashed dotted lines represent the mean and median of the instantaneous ram pressure. The solid vertical line is the maximum ram pressure we expect a galaxy in the Coma cluster could experience.
\label{fig:Pram_Hist_Median}}
\end{figure}

Taking the median instead of the mean of the average sSFRs does not produce significantly different results. The overall trend with the Coma sample still decreasing more rapidly as a function of galactocentric radius than the control is still preserved (Fig.~\ref{fig:Model_Median}), and the fitted ram pressure 
is roughly the same as the one produced by fitting to the mean (Fig.~\ref{fig:Model_Median}, Fig.~\ref{fig:Corner_Plot_Median}, Fig.~\ref{fig:Pram_Hist_Median}). The fitted ram pressure is $1.7\,_{-0.9}^{+0.7} \times 10^{-11}\,\mathrm{g\,cm^{-1}\,s^{-2}}$ and a $\Delta \mathrm{sSFR}$ of $-0.41_{-0.16}^{+0.15}$.


\bibliography{sample631}{}
\bibliographystyle{aasjournal}

\end{document}